\documentclass[a4paper,fleqn,usenatbib,useAMS]{mnras}

\usepackage{graphicx}	
\usepackage{amsmath}	
\usepackage{amssymb}	
\usepackage{multicol}        
\usepackage{bm}		
\usepackage{pdflscape}	
\usepackage{color}

\usepackage[T1]{fontenc}
\usepackage{ae,aecompl}

\usepackage{newtxtext,newtxmath}

\title[Rayleigh-Taylor instability in relativistic jets]{Rayleigh-Taylor Instability in Two-Component Relativistic Jets}

\author[K. Toma, S. S. Komissarov \& O. Porth]{Kenji Toma$^{1,2}$\thanks{Contact e-mail: \href{mailto:toma@astr.tohoku.ac.jp}{toma@astr.tohoku.ac.jp}}, Serguei S. Komissarov$^{3}$ and Oliver Porth$^{4}$
\\
$^{1}$Frontier Research Institute for Interdisciplinary Sciences, Tohoku University, Sendai 980-8578, Japan \\
$^{2}$Astronomical Institute, Tohoku University, Sendai 980-8578, Japan \\
$^{3}$School of Mathematics, University of Leeds, Leeds, LS2 9JT, UK \\
$^{4}$Institute for Theoretical Physics, Max-von-Laue-Str. 1, 60438, Frankfurt am Main, Germany \\
}

\begin{document}
\label{firstpage}
\pagerange{\pageref{firstpage}--\pageref{lastpage}}
\maketitle

\begin{abstract}
Relativistic jets associated with active galactic nuclei and gamma-ray bursts propagate over huge distances without significant loss of momentum.  At the same time they are bright emitters, which is indicative of strong energy dissipation. This points towards a mechanism of internal dissipation which does not result in a global disruption of the flow.  One possibility is internal shocks and another one is turbulence driven by local instabilities. Such instabilities can be triggered when a freely expanding jet is reconfined by either the cocoon or external gas pressure. In this paper we study the dynamics of two-component spine-sheath hydrodynamic jets coming into pressure 
equilibrium with external gas using 2D computer simulations.  We find that the jet oscillations lead to a rapid onset of Rayleigh-Taylor-type instabilities, which results in additional internal dissipation and mixing of the jet components.   Although slightly different in details, this outcome holds both for the heavy-spine-light-sheath and light-spine-heavy-sheath configurations.  The results may provide an explanation to the spatial flaring observed in some AGN jets on kpc-scales.
\end{abstract}





\section{Introduction}
\label{sec:intro}

Relativistic collimated outflows (jets) is  a relatively common astrophysical phenomenon. In spite of the significant progress in recent years, there are still many unresolved key issues concerning their production, acceleration, mass-loading, collimation, and energy dissipation/emission mechanisms.   
In particular, these outflows are widely believed to be driven by black-hole-accretion-disk systems and hence can have components associated both with the black hole outflow powered by the Blandford-Znajek mechanism \citep{blandford77} and the collimated disk wind powered by the Blandford-Payne mechanism \citep{BP82}. The relative importance of these components remains unclear and often the astrophysical jets are studied using simple one-component models. On the other hand, during the past decade computer simulations of astrophysical flows in general and relativistic jets in particular reached  high levels of sophistication allowing to explore much more realistic models.   

In stark contrast to terrestrial and laboratory jets, the astrophysical jets somehow manage to penetrate vast space without suffering destructive instabilities. It is not unusual for jets of active galactic nuclei (AGN) to be traced up to distances exceeding the initial jet radius by a billion times.  On the other hand, their observed emission indicates local dissipation of their kinetic or magnetic energy. One popular physical mechanism of such dissipation involves shock waves. Alternatively, the dissipation may be triggered by instabilities.  
 
So far the studies of jet stability have been focused on the Kelvin-Helmholtz and current-driven magnetic instabilities of cylindrical flows  \citep[e.g.][]{lyubarsky99,appl00,baty02,hardee03,nakamura04,mizuno07,mizuno12,kim17}. The cylindrical configuration was selected mainly to allow for closed-form analytic solutions but also  to simplify computer simulations. In contrast, the astrophysical jets are normally collimated only with an opening angle of few degrees and hence exhibit strong lateral expansion. It has been suggested that this expansion plays a crucial role in ensuring the observed stability of these jets via hindering causal communication across them \citep[e.g.][]{moll08,porth15}.  In a zone with a relatively flat external pressure distribution, the causal communication may eventually get restored via the so-called reconfinement process, which involves a strong conical shock\footnote{A shock which results in overall increase of the jet collimation is often called a {\it recollimation shock}. Such a shock does not necessarily restore the causal connectivity across the jet and may even never reach its axis \citep[e.g.][]{KBB12}. When it does, we agree to call it a {\it reconfinement shock}.} driven into the jet by the external gas pressure. The steady-state solutions of reconfined jets are characterised by an overall increase of the jet collimation accompanied by decaying radial oscillations \citep[e.g.][]{Sanders83,FW85,komissarov15}.

The accelerated motion of shocked jet plasma during the reconfinement process is different from the constant poloidal velocity motion of cylindrical jets and this may lead to instabilities absent in purely cylindrical configurations. For example, it has been shown that the oscillations of initially over-pressured (compared to the external gas) 2D jets with translational symmetry may lead to an onset of  Rayleigh-Taylor-type (RT-type) instabilities  \citep[][hereafter MM13]{matsumoto13}.  Indeed, in the accelerated frame moving up and down with the jet-external gas interface both fluids are subject to a non-inertial gravity force.   MM13 argued that this 2D problem captured the essence of the spatial oscillations in a steady-state recollimated jet, where the centrifugal force plays the role of gravity\footnote{The rotation-induced instability of two-component relativistic jets was studied in \citet{meliani07,meliani09} and recently by \citet{millas17} who extended this study to magnetised jets. While dynamically important close to the central engine, the rotation is expected to be too week at the kpc-scales where the AGN jets become reconfined.}.   

Based on the results of their 2D simulations, MM13 proposed the following empirical condition for the instability 
\begin{equation}
\eta >1 \quad\mbox{where}\quad
\eta = \frac{\rho_{{\rm j}} h_{{\rm j}} \Gamma_{{\rm j}}^2}{\rho_{\rm ext} h_{\rm ext}} \,.
\label{eq:RTcondition}
\end{equation}
In this expression  $\rho_{\rm j}$, $h_{\rm j}$, and $\Gamma_{\rm j}$ are the initial jet parameters (rest mass density, specific enthalpy and Lorentz factor respectively) and $\rho_{\rm ext}$  and $h_{\rm ext}$ are the corresponding parameters of the external gas.   For a high Lorentz factor and either cold or relativistically hot external medium, $\eta$  is simply the ratio of mass-energy densities, implying that this criterion is a straight-forward  generalisation of the original RT instability condition. Indeed, as the jet expansion caused by the initial pressure imbalance slows down, the acceleration vector in the frame of the contact points outwards and the configuration appears RT-unstable if the jet is heavier than the external gas, that is $\eta>1$. MM13 argued that in the case of reconfined jets this phase corresponds to the collimation episodes of the jet oscillation.

MM13 argued that the instability criterion (\ref{eq:RTcondition}) can be met by jets which are surrounded by light cocoons filled with jet plasma processed at the jet termination shock.  However, not all AGN jets show evidence of such cocoons and could be surrounded  by the relatively dense interstellar gas instead.  Based on the X-ray emission of this gas one can deduce its pressure and it turns out to be sufficient to force reconfinement of jets with power $L \le 10^{44}\;{\rm erg\;{\rm s}^{-1}} $  on the scales $\le 1\,$kpc \citep[e.g.][]{porth15}.   Such power is typical of the Fanaroff-Riley type I jets  \citep{FR74} and hence this case is of astrophysical interest.  Using the typical parameters of the X-ray coronas of elliptical galaxies we find that at such scales 
\begin{equation}
  \eta  \simeq 10^{-3}\, {L_{\rm j,44}} \, z_{\rm kpc}^{-2}\, \theta_{\rm j,-2}^{-2}\, n_{\rm ext}^{-1},
\end{equation}
where $L_{\rm j,44}$ is the jet power in the units of $10^{44}\,{\rm erg\, s^{-1}}$, $z_{\rm kpc}$ is the distance from the galactic centre in kpc, $\theta_{\rm j,-2}$ is the jet opening angle in the units of $10^{-2}\,$rad and $n_{\rm ext}$ is the number density of the coronal gas in CGS units.  Hence, according to the MM13 criterion such ``naked'' jets should be RT-stable.  

However, structured naked jets may still develop internal RT-type instabilities. For example, a heavy-spine-light-sheath (HSLS) jet has an interior structure which is similar to the heavy-jet-light-cocoon (HJLC) configuration of MM13 and hence  should become RT-unstable in the reconfinement zone. Light-spine-heavy-sheath (LSHS) jets have the inverse density structure, which by analogy with the light-jet-heavy-cocoon (LJHC) configuration seems to indicate that the spine-sheath interface can be RT-stable.  However, this analogy is not exact  and the interface could turn unstable during the de-collimation episodes of the reconfinement process where the acceleration vector changes its direction.  As the oscillations of reconfined jets involve shock waves, the Richtmyer-Meshkov instability \citep[RMI,][]{RDR,EEM}, which can be described as an impulsive version of RT instability, may also play a role (MM13).   

In this paper, we report the results of the very first study into the dynamics of unmagnetised structured relativistic jets during reconfinement. In this study, we  employed the same numerical approach as MM13 and considered both HSLS and LSHS structured configurations.  We have found that for both the jet types the contact discontinuity between the spine and the sheath is subject to RT-type instabilities which leads to efficient mixing and dissipation inside the jet on the reconfinement scale.

\begin{figure}
  \begin{center}
    \includegraphics[width=\columnwidth]{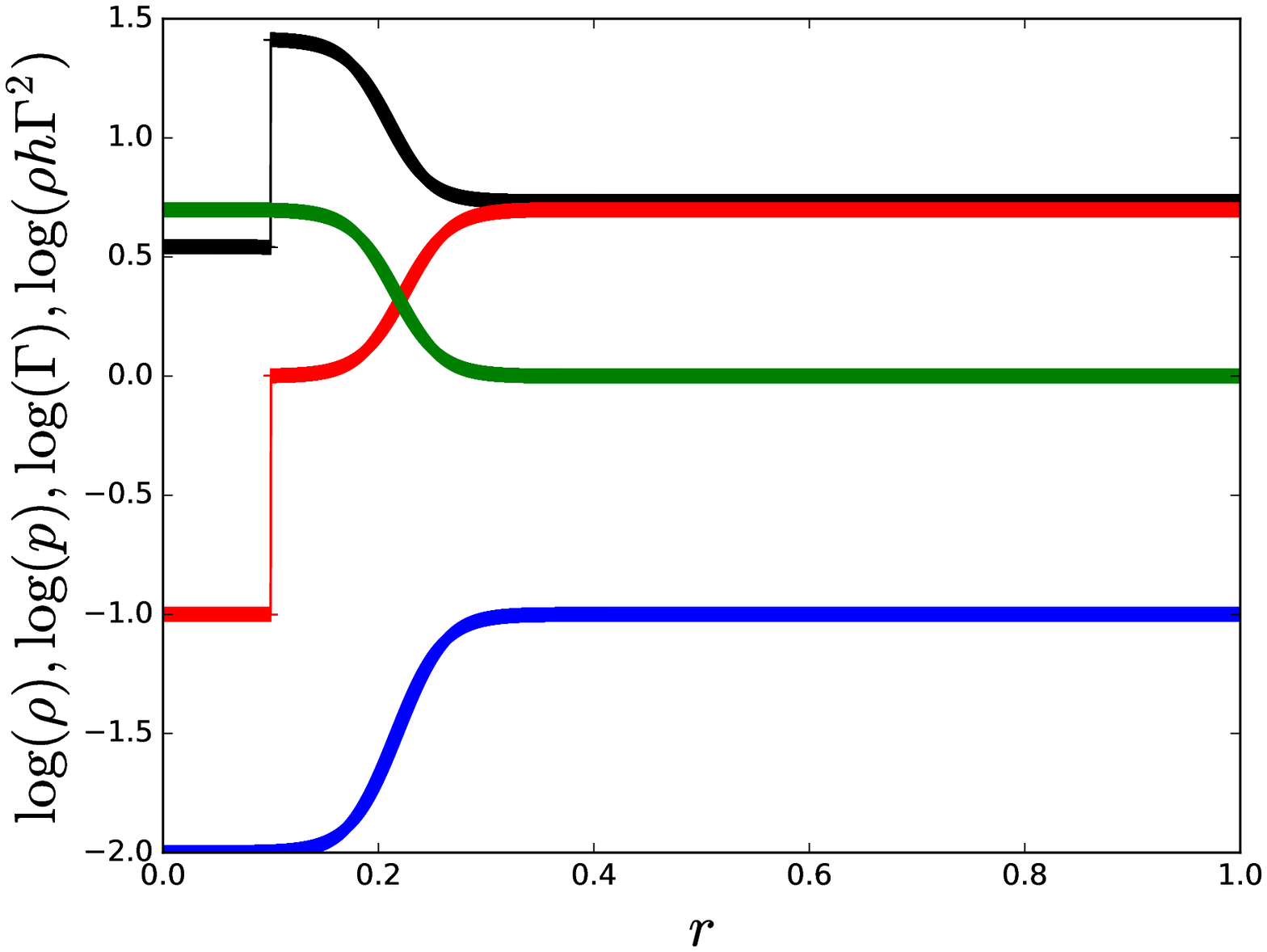}
  \end{center}
 \caption{The initial conditions of the LSHS problem.  The curves show $\log \rho$ ({\it red}), $\log p$ ({\it blue}), $\log \Gamma$ ({\it green}) and $\log \rho h \Gamma^2$  ({\it black}).}
 \label{fig:lsj_t0}
\end{figure}

\begin{figure*}
\begin{tabular}{c}
  \begin{minipage}{0.35\hsize}
    \includegraphics[scale=0.7]{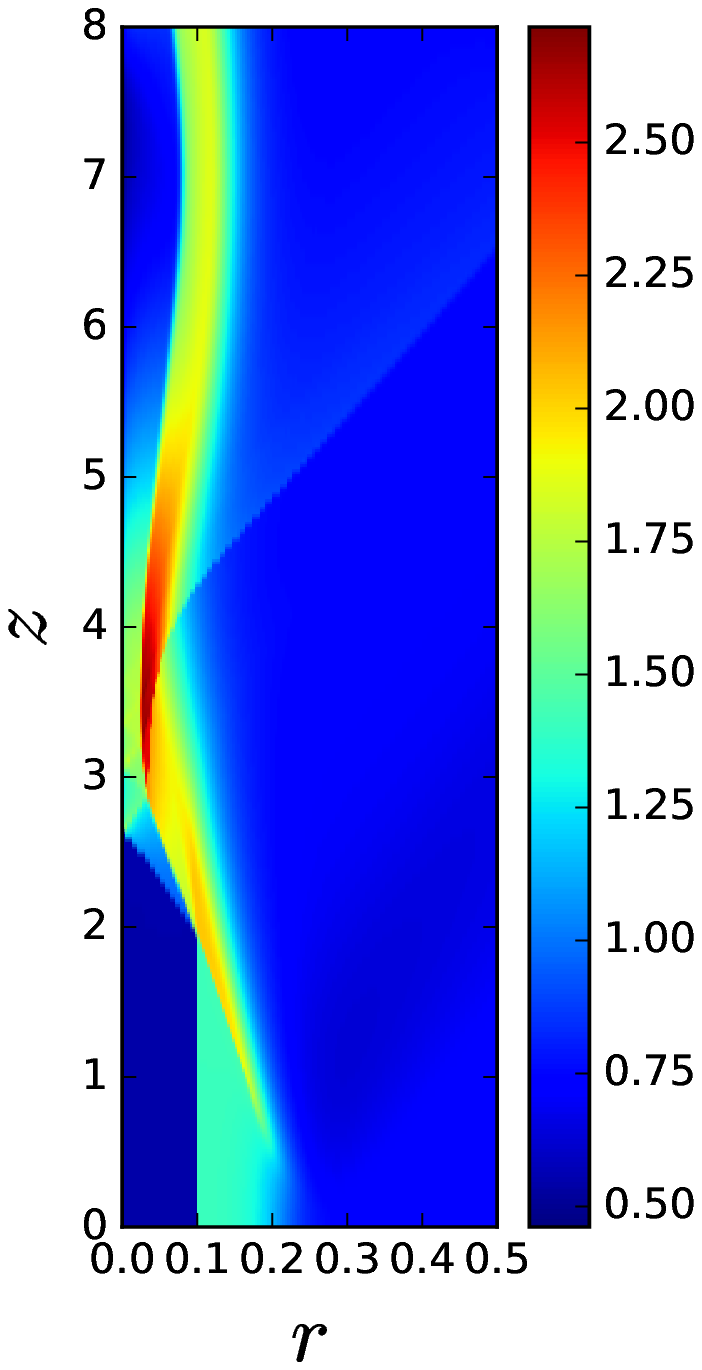}
  \end{minipage}
  \begin{minipage}{0.65\hsize}
    \begin{tabular}{c}
      \begin{minipage}{0.5\hsize}
        \begin{center}
          \includegraphics[scale=0.4]{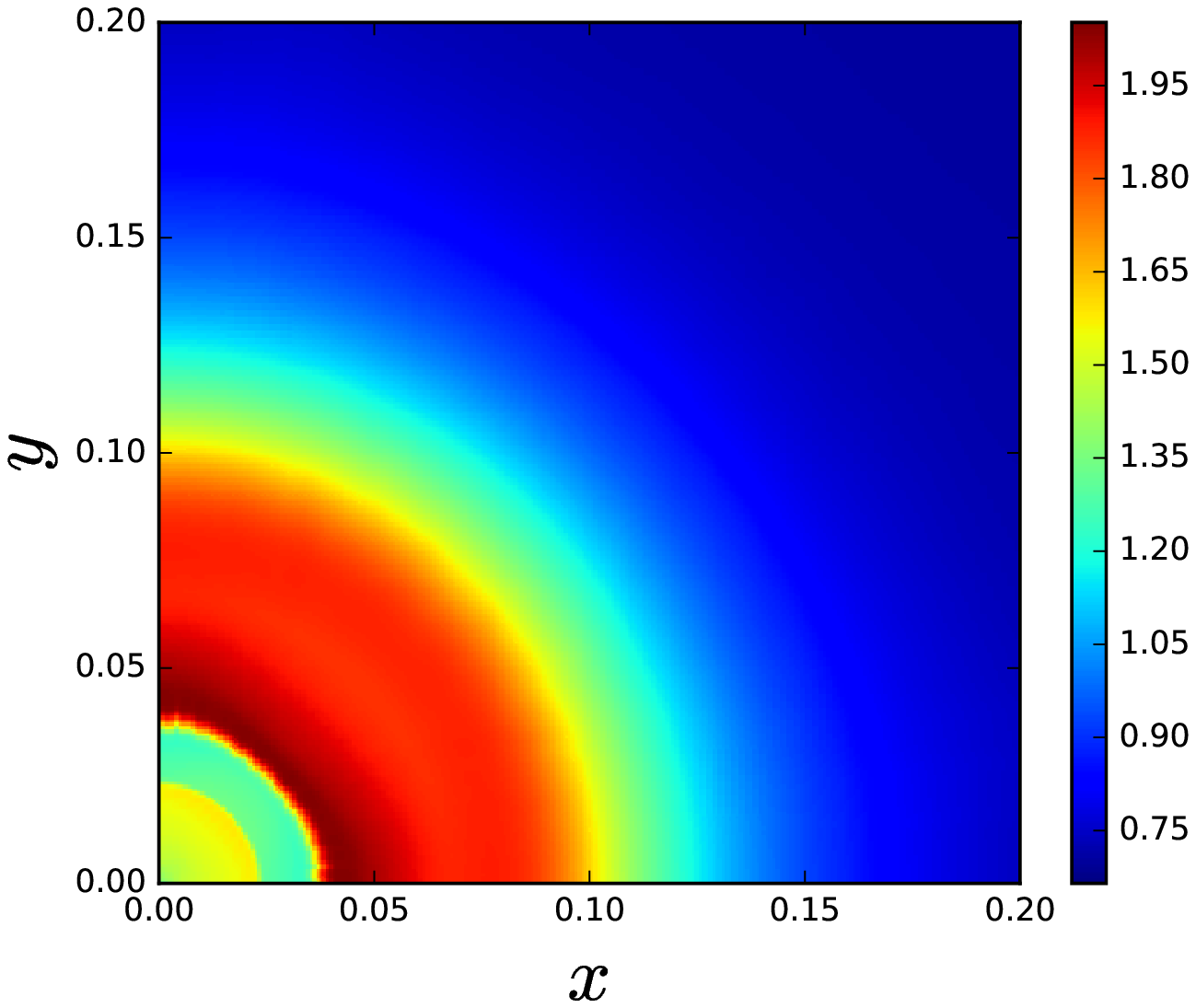}
        \end{center}
      \end{minipage}
      \begin{minipage}{0.5\hsize}
        \begin{center}
          \includegraphics[scale=0.4]{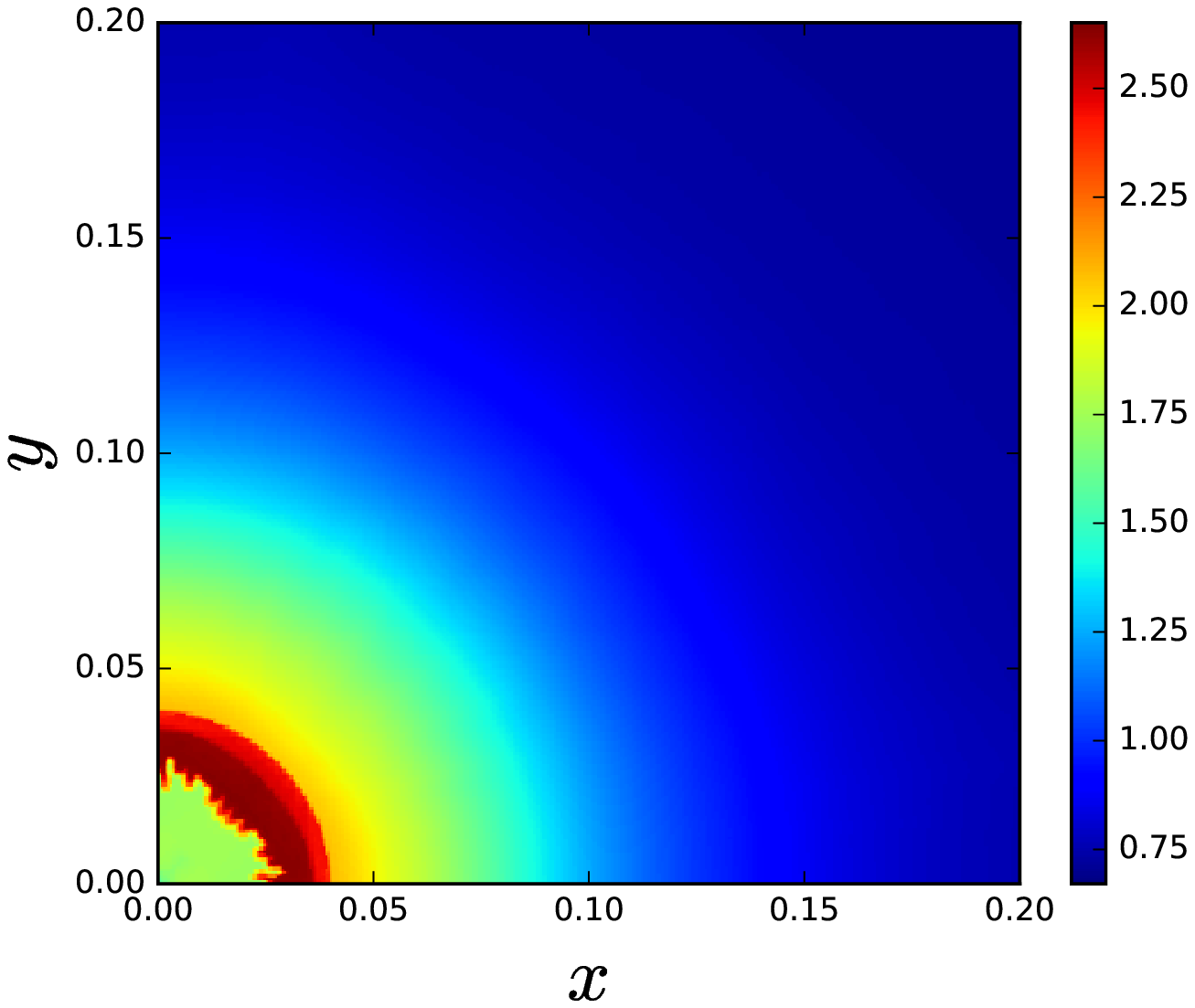}
        \end{center}
      \end{minipage}
    \end{tabular}
    \begin{tabular}{c}
      \begin{minipage}{0.5\hsize}
        \begin{center}
          \includegraphics[scale=0.4]{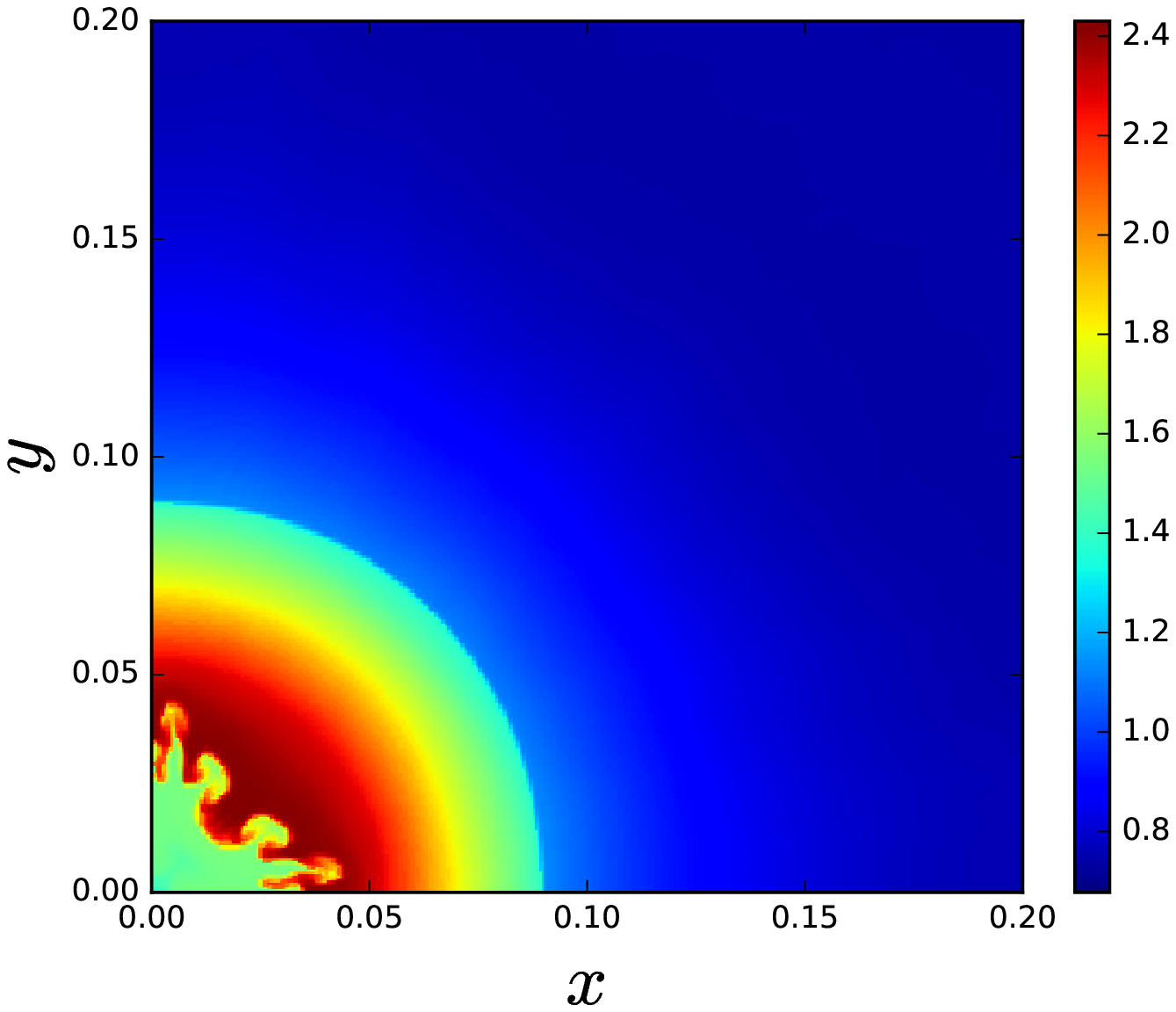}
        \end{center}
      \end{minipage}
      \begin{minipage}{0.5\hsize}
        \begin{center}
          \includegraphics[scale=0.4]{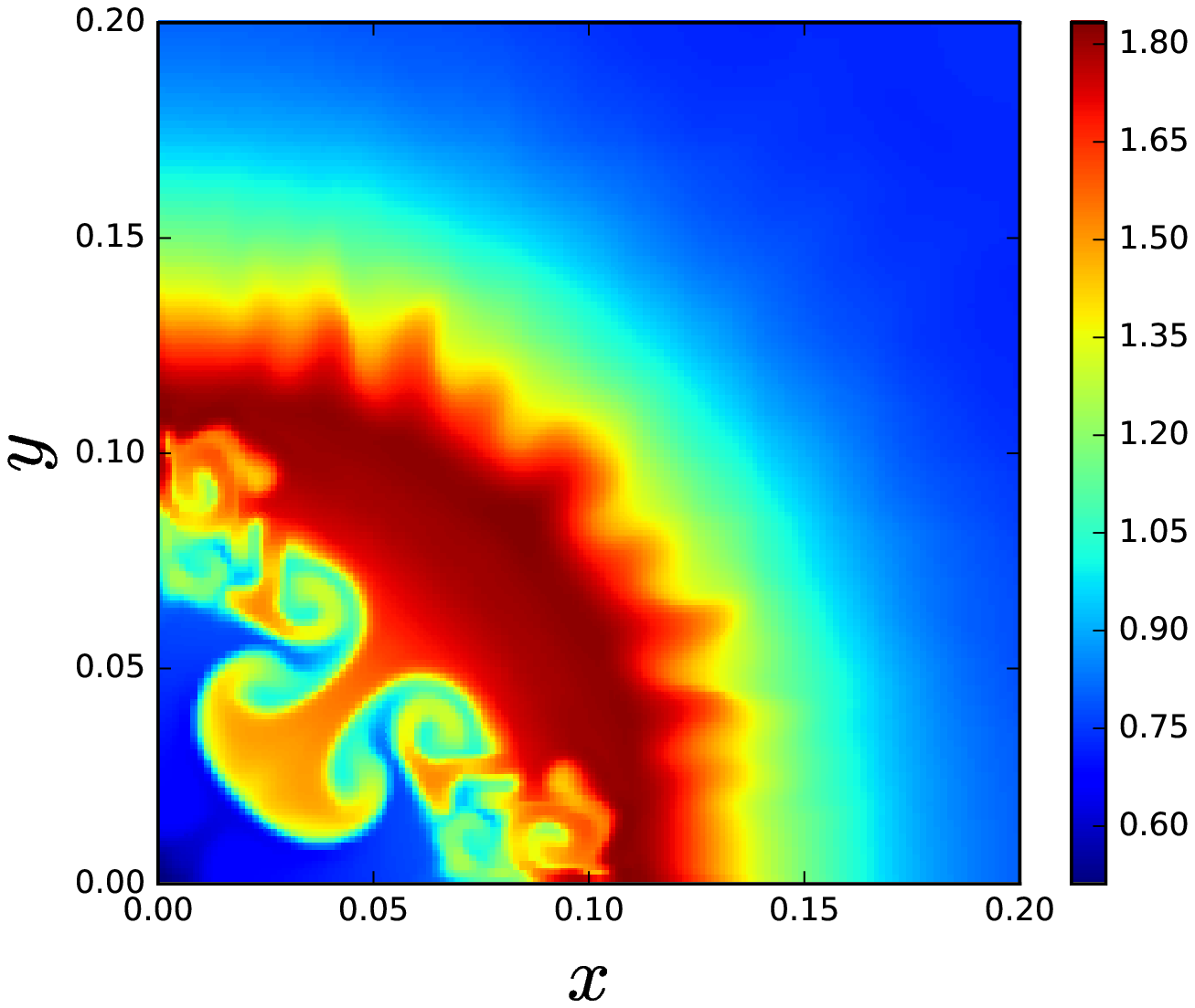}
        \end{center}
      \end{minipage}
    \end{tabular}
  \end{minipage}
\end{tabular}
\caption{Left: the steady-state solution for the LSHS jet based on 1D time-dependent simulations.   Right: the transverse structure of the jet with the same initial condition as the Left panel
at  $z=2.6$ ({\it top-left}), $z=3.2$ ({\it top-right}), $z=4$ ({\it bottom-left}), and $z=7$ ({\it bottom-right}) based on 2D time-dependent simulations. The parameter shown is the effective inertia $\log(\rho h \Gamma^2)$.}
\label{fig:lsj}
\end{figure*}

\section{Method}
\label{sec:setup}

Our approach is first to obtain a numerical stationary axisymmetric jet solution and then to explore its stability using 2D time-dependent simulations.  For the first step we use a novel numerical method of finding approximate steady-state relativistic jet solutions, which was originally proposed by  \citet{matsumoto12} and then rigorously developed by  \citet[][hereafter KPL15]{komissarov15}. In this method, the steady-state jet structure is reconstructed via 1D time-dependent simulations in cylindrical geometry. This approach is based on the similarity between $\partial/\partial t$ in the time-dependent equations and $\partial/\partial z$ in the steady-state equations for narrow jets with the axial velocity $v_z\approx c$.  Once the time-dependent 1D solution is found, the steady-state solution is obtained via the substitution $t=z/c$. Thus the time evolution of the 1D solution corresponds to the spatial evolution along the jet axis of the steady-state axisymmetric  solution.   In this approach, the treatment of external gas is a rather delicate issue. Indeed, in the steady-state problem the static external gas remains unaffected by the jet, whereas in the 1D problem the jet may emit waves. KPL15 describe a special procedure for updating the external gas parameters designed to minimise this emission. Alternatively, one may simply set the external gas density to a very low value.  For technical reasons, the latter approach is adopted here.

For the stability study we adopt the same approach as MM13, who numerically integrated the time-dependent equations of relativistic gas dynamics in cylindrical geometry with the imposed symmetry condition $\partial/\partial z=0$, where the z coordinate is measured along 
the jet axis\footnote{Hence we are looking for solutions which depend only on the cylindrical coordinates $r$ and $\phi$.}.  They considered initially over-pressured jets and studied their oscillations triggered by this initial lack of balance.  In fact, this is very similar to what we do in our 1D simulations, with just one additional degree of freedom, suggesting that this way we study the spatial growth of modes orthogonal to the jet axis. That is, we deal with the perturbations of the form  $f(r) \exp(i(\omega t + m\phi + k z))$ with $Im(\omega)=Re(k)=0$ but allowing $Im(k)\not=0$.  As a side-effect, this approach suppresses all Kelvin-Helmholtz modes, which require $Re(k)\not=0$, thus allowing us to focus on the evolution of RT and RM type instabilities in isolation. 

This 2D approach shares not only the advantages but also the limitations of the 1D method of KPL15. In particular, it is not suitable to deal with the case of dense external gas and hence naked AGN jets. In this case the jet oscillations become over-damped via the wave emission. Hence we are forced to deal with the case of light external gas.  Since we are interested in the stability of the spine-sheath interface this does not seem to be much of a problem. However, for a light external gas $\eta>1$ and the jet boundary is also RT-unstable. In the LSHS case, this instability develops rapidly during the initial expansion of an over-pressured (or a freely-expanding) jet and changes the flow dramatically before the spine-sheath interface becomes unstable at the end of the first contraction phase.  This can be helped by considering under-pressured jets with $v_r=0$ at the ``inlet'' as this  allows to avoid the initial expansion phase.  The reconfinement shock is still driven into the jet, and the radial oscillations are still triggered downstream.

The simulations are carried out using MPI-AMRVAC code which utilises HLLC Riemann solver, second-order spatial TVD reconstruction and a three-step Runge-Kutta time integrator \citep{keppens12,porth14}\footnote{https://gitlab.com/mpi-amrvac/amrvac}. The equation of state is $h = 1+[\gamma/(\gamma-1)]p/\rho$ with the adiabatic index $\gamma=4/3$.

\section{Results}

\subsection{LSHS jets}
\label{sec:lsj}

Here we present the results for the model with inlet flow parameters   
\begin{eqnarray}
  && \rho = \begin{cases}
    0.1 \quad\mbox{for}\quad  r<0.1,\\
    3 + 2 \tanh[(r-0.24)/0.04] \quad\mbox{for}\quad r>0.1, 
  \end{cases} \nonumber \label{eq:lsj} \\
  && p = 0.055 + 0.045 \tanh[(r-0.24)/0.04], \\
  && \Gamma = 3 - 2 \tanh[(r-0.2)/0.04]. \nonumber
\end{eqnarray}
These are illustrated in  Figure~\ref{fig:lsj_t0} which also shows the effective inertia $\rho h \Gamma^2$. One can see that the light spine extends up to  $r\approx 0.1$ and the heavy sheath occupies $0.1 < r \lesssim 0.3$.  The effective inertia ratio between the two components is $\approx 7$.
The spine Lorentz factor $\Gamma_{\rm j} \approx 5$ and it slowly reduces to unity in the spine. The jet pressure is $\approx 10$ times lower than that of the external medium. The computational domain spans $0<r<1$. 

The steady-state solution is illustrated in the left panel of Figure~\ref{fig:lsj}, which shows the distribution of $\log(\rho h \Gamma^2)$. One can see that the external pressure drives a reconfinement shock which first crosses the sheath and then the spine of the jet. It gets reflected off the axis at $z\approx 2.6$, turning into a de-collimation shock, and then reaches the sheath at $z\approx 2.9$. At this point it breaks and then runs at a noticeably more acute angle to the jet axis, which reflect the higher ram pressure of the sheath. The shocked sheath forms a dense shell about the jet axis between $z=3$ and $z=5$.
At $z\approx 4$ a compression wave detaches from the jet and then reaches the plot's right boundary $r = 0.5$ at $z\approx 6.5$ -- this is an example of the artificial wave emission typical for the 1D method. The wave is rather week though.

For the time-dependent simulations, we use a Cartesian domain with  $-1 < x < 1$ and $-1 < y < 1$ and employ AMR with five levels of refinement and the base level of  $160 \times 160$ cells. For the refinement criterion we employ the phenomenological L\"{o}hner's error estimator \citep{lohner87} applied to the fluid density and vertical momentum with equal weights. We use the same initial solution as above and do not introduce any perturbations. However, the projection of the initial axisymmetric solution on the Cartesian grids automatically generates grid-size fluctuations of the physical quantities.
The results are illustrated in the four right panels of Figure~\ref{fig:lsj} which show the distribution of  $\log(\rho h \Gamma^2)$ at times corresponding to the distances of $z=2.6$, $3.2$, $4$, and $7$ along the steady-state solution.   At $z=2.6$ the reconfinement shock moves away from the jet axis after being reflected earlier. It has not reached yet the spine-sheath interface, which shows signs of a small-amplitude disturbance.  At $z=3.2$ the shock has passed the interface which now exhibits a strong corrugation.  Further downstream the corrugation takes the form of heavy fingers reaching inside the light spine which eventually develop the mushroom shape so typical of the RT-type instabilities.   Overall, the observed evolution is consistent with the impulsive RM instability. 
  
Interestingly, at $z=7$ there are signs of another instability developing in the outer layer of the sheath.  This layer is not uniform, with the effective inertia decreasing outwards.  As the streamlines of the flow are convex the conditions in the layer are favourable to the RT instability and this is what is observed.  

In addition to this model, we have also studied a number of other models with a lower central density of the spine. They all show a similar evolution with clear signatures of  the RT and RM instabilities.

\subsection{HSLS jets}
\label{sec:hsj}

\begin{figure}
  \begin{center}
    \includegraphics[width=\columnwidth]{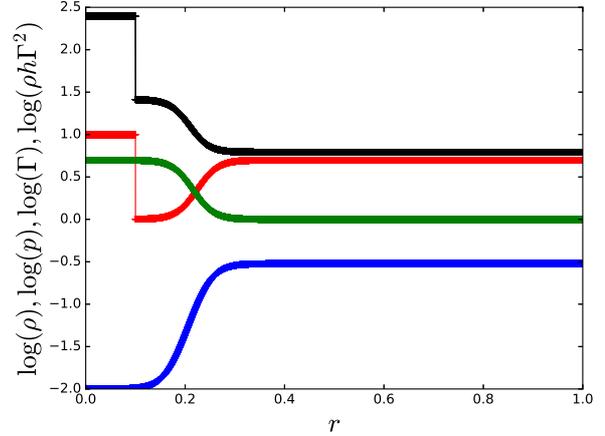}
  \end{center}
    \caption{The initial conditions of the HSLS problem.  The curves show $\log \rho$ ({\it red}), $\log p$ ({\it blue}), $\log \Gamma$ ({\it green}) and $\log \rho h \Gamma^2$  ({\it black}).}
 \label{fig:hsj_t0}
\end{figure}

\begin{figure*}
\begin{tabular}{c}
  \begin{minipage}{0.35\hsize}
    \includegraphics[scale=0.7]{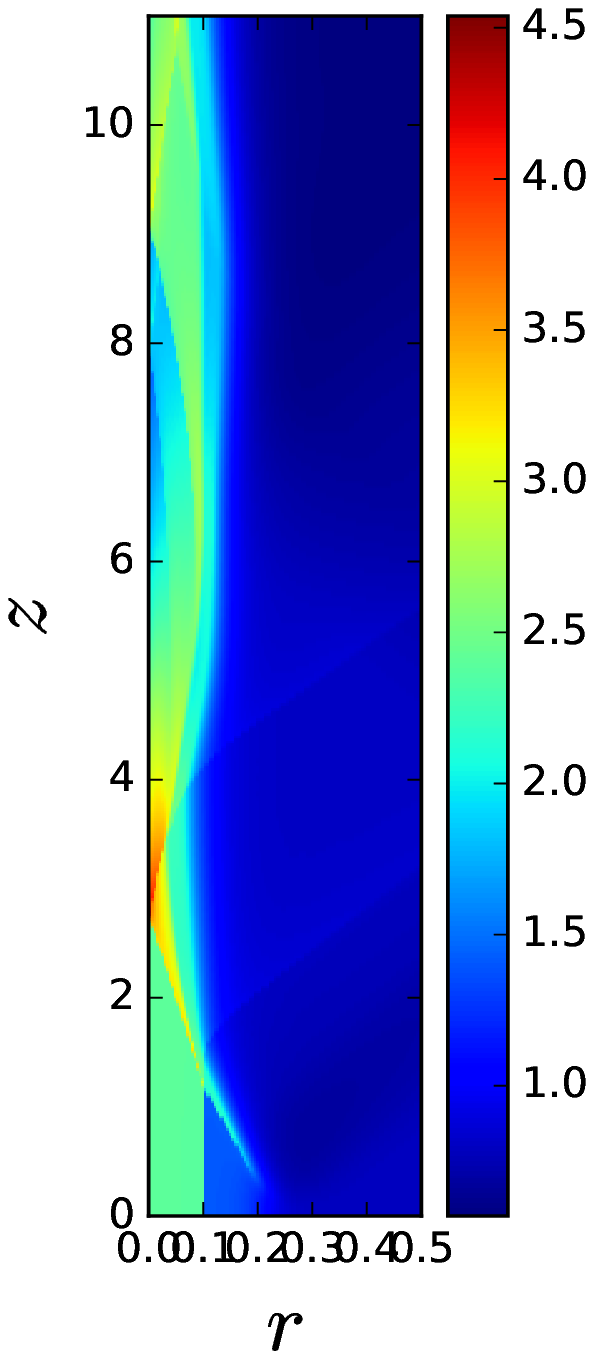}
  \end{minipage}
  \begin{minipage}{0.65\hsize}
    \begin{tabular}{c}
      \begin{minipage}{0.5\hsize}
        \begin{center}
          \includegraphics[scale=0.4]{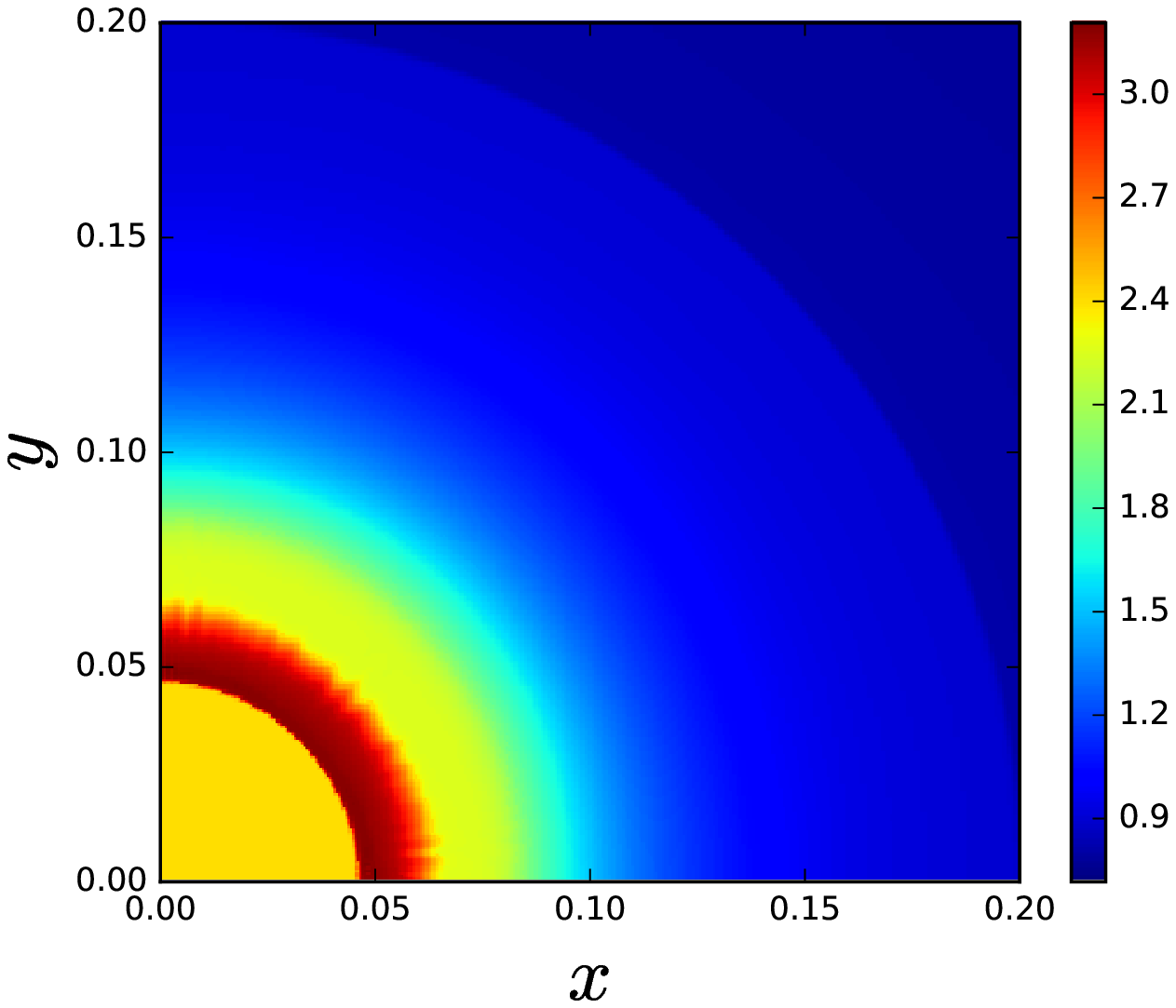}
        \end{center}
      \end{minipage}
      \begin{minipage}{0.5\hsize}
        \begin{center}
          \includegraphics[scale=0.4]{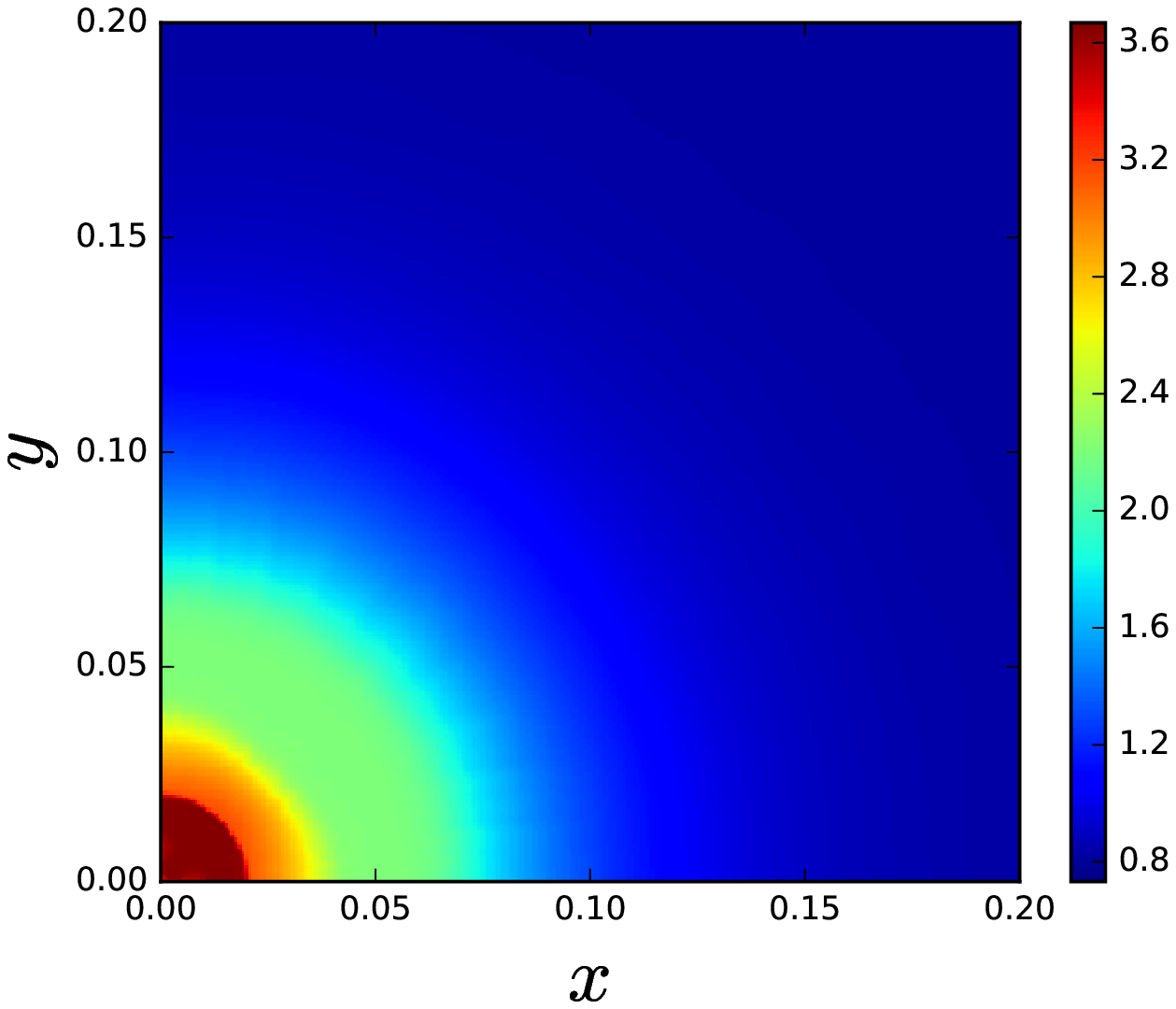}
        \end{center}
      \end{minipage}
    \end{tabular}
    \begin{tabular}{c}
      \begin{minipage}{0.5\hsize}
        \begin{center}
          \includegraphics[scale=0.4]{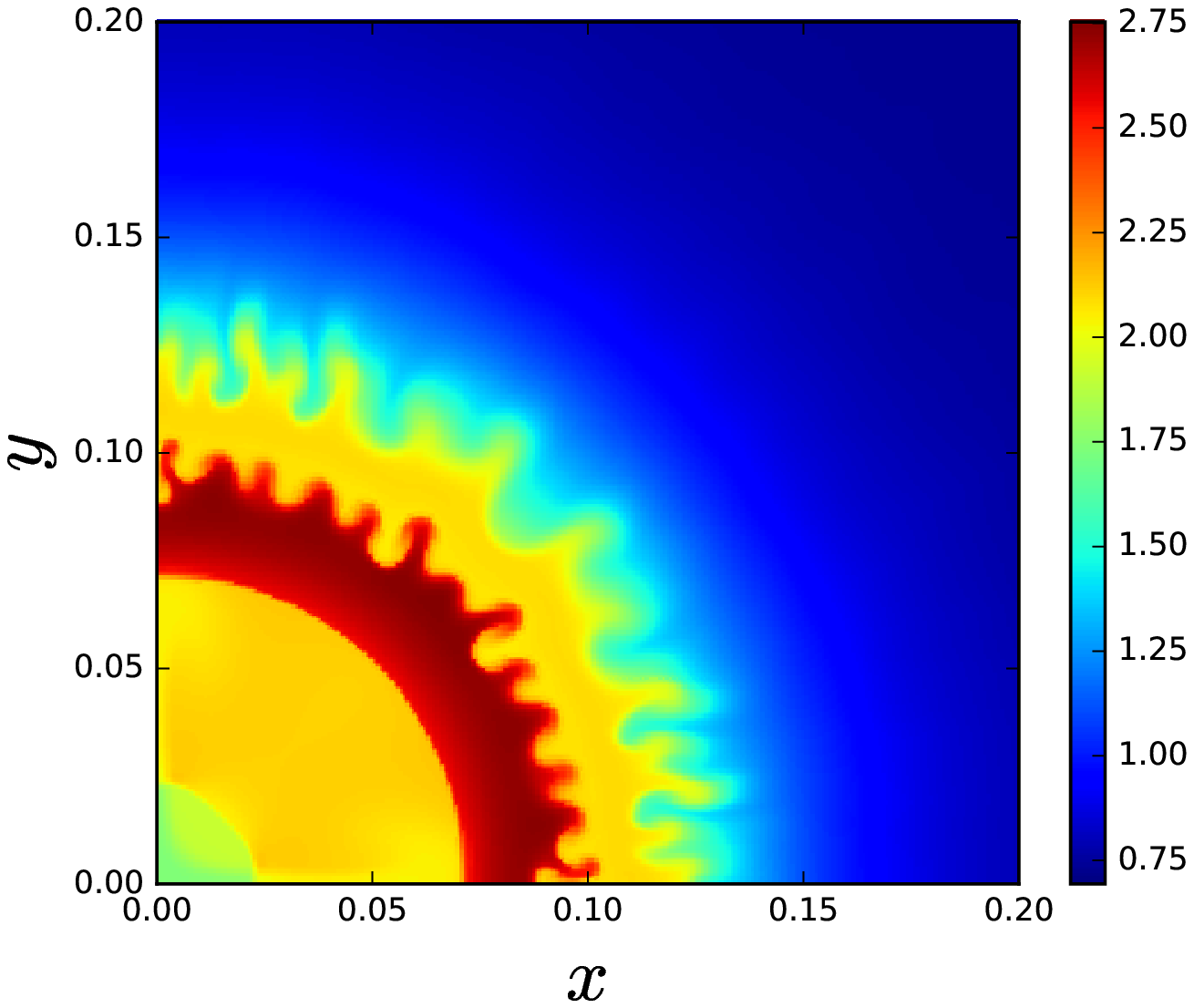}
        \end{center}
      \end{minipage}
      \begin{minipage}{0.5\hsize}
        \begin{center}
          \includegraphics[scale=0.4]{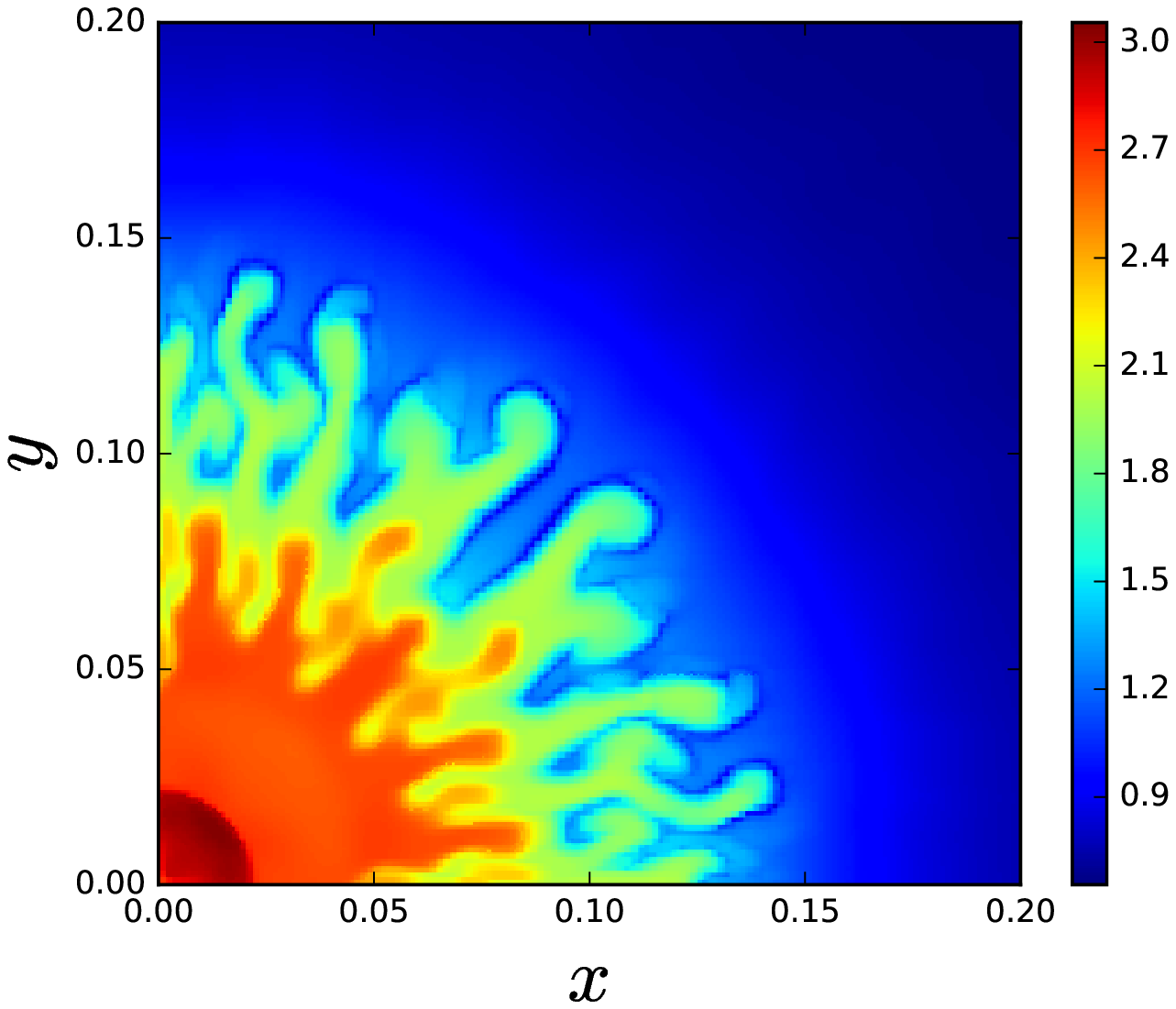}
        \end{center}
      \end{minipage}
    \end{tabular}
  \end{minipage}
\end{tabular}
\caption{Left: the steady-state solution for the HSLS jet based on 1D time-dependent simulations.   Right: the transverse structure of the jet with the same initial condition as the Left panel
at   $z=2$ ({\it top-left}), $z=3.2$ ({\it top-right}), $z=7$ ({\it bottom-left}), and $z=9.5$ ({\it bottom-right}) based on 2D time-dependent simulations. The parameter shown is the effective inertia $\log(\rho h \Gamma^2)$.}
 \label{fig:hsj}
\end{figure*}

Here we present the results for the model with inlet flow parameters   
\begin{eqnarray}
  && \rho = \begin{cases}
    10.0 \quad\mbox{for}\quad  r<0.1,\\
    3 + 2 \tanh[(r-0.24)/0.04] \quad\mbox{for}\quad  r>0.1, 
  \end{cases} \nonumber \label{eq:hsj} \\
  && p = 0.155 + 0.145 \tanh[(r-0.24)/0.04], \\
  && \Gamma = 3 - 2 \tanh[(r-0.2)/0.04]. \nonumber
\end{eqnarray}
These distributions are shown in Figure~\ref{fig:hsj_t0}. The corresponding steady-state solution is illustrated in the left panel of Figure~\ref{fig:hsj} which shows the distribution of the effective inertia, $\log(\rho h \Gamma^2)$.   The jet initially contracts in a similar manner to the LSHS jet, then  bounces at $z \sim 3$ and develops oscillations. 

For the time-dependent simulations we used the grid with the same number of AMR levels as in the LSHS case. The results are illustrated in the right panels of Figure~\ref{fig:hsj}, which shows the solution at the times corresponding to the locations with the distances of $z=2, 3.2, 7$, and $9.5$ in the steady-state model.  At $z=2$ the reconfinement shock has crossed the spine-sheath interface, which now shows small-amplitude disturbance. Its amplitude remains small all the way to the bounce point. After the bounce, the solution clearly exhibits two unstable zones with outwards reaching  fingers of heavy gas.  The inner one corresponds to the RM-unstable spine-sheath interface and the outer one to the RT-unstable outer layer of the sheath where the effective inertia gradually decreases outwards.  Hence our results confirm and strengthen the MM13 conclusions by capturing the case where the discontinuity is replaced by a smooth layer.

\section{Conclusion and Dicussion}
\label{sec:conclusion}

Along their length, AGN jets (and many other astrophysical jets) almost inevitably experience reconfinement  by the pressure of either the surrounding interstellar gas  or their own cocoons.  We have studied the dynamics of reconfinement in the case of spine-sheath hydrodynamic jets moving through uniform external medium. We find that the reconfinement gives rise to RT-type instabilities at the contact discontinuity between the spine and the sheath and results in mixing of the two components. This occurs both in the light-spine-heavy-sheath and heavy-spine-light-sheath cases, because each reconfinement episode is followed by a de-collimation, which involves switching the effective gravity direction. The instability grows on the reconfinement scale, soon after (or just before) the reconfinement shock reaches the jet axis in the light-spine-heavy-sheath (heavy-spine-light-sheath) case. Our results also show that the instability growth in layers with initially smooth variation of effective inertia is as fast as at contact surfaces with discontinuous effective inertia. This allows us to conclude that the reconfinement is accompanied by efficient mixing of any parts with strong radial variation of inertial density in hydrodynamic jets.

It is natural to assume that all astrophysical jets are born structured, due to the nature of their central engines. In particular, the  accreting central engines  may have a spine-sheath structure, where the spine is connected to the central compact object and the sheath to the accretion disc. For this reason, we conclude that reconfinement of all weakly-magnetised astrophysical jets should lead to rapid onset of RT-type instabilities and efficient mixing. The spine-sheath jet structure has been also discussed in the observational context of AGN jets, e.g. to explain the high-resolution images and polarimetric data of radio jets \citep[e.g.][]{LB14,gabuzda13} as well as the broadband spectra of blazer jets \citep{ghisellini05}.

Fluid instabilities normally result in enhanced turbulence and dissipation of both kinetic and magnetic energy. In the AGN context, such conditions are favourable for nonthermal particle acceleration via second-order Fermi mechiansm and magnetic reconnection. This could be the reason behind the observed flaring of the Fanaroff-Riley type I jets on kpc-scales \citep{LB14}, scales where these low power jets are expected to get reconfined by the thermal pressure of the galactic X-ray coronas \citep{porth15}. The shock acceleration mechanism has become rather less attractive since the recent particle-in-cell simulations have shown that it is not activated unless the magnetic to kinetic energy density ratio $\sigma \lesssim 10^{-3}$ \citep{sironi11}.

A possible connection between the reconfinement and jet flaring has been already discussed in application to the prominent M87 jet \citep{AN12}.  In this context, it is tempting to interpret the conical opening of the M87 jet after encountering the HST-1 knot as a consequence of the free-streaming motion performed by the heads of the internal RT fingers.  

In our study, we only considered unmagnetised jets. However, the central engines of AGN jets are almost certainly magnetic and the jets  are expected to remain magnetically dominated up to pc-scales \citep{komissarov07}.  Strong magnetic field may stabilise the flow against RT instability \citep[e.g.][]{millas17}  but bring into play current-driven instabilities. This issue has to be explored in future studies of the reconfinement process. Even when the magnetic field is too weak to suppress RT instability, it has to be taken into account when modelling jets' non-thermal emission.
Finally, full 3D simulations are needed to overcome the obvious limitations of our 2D simulations, to treat properly the interaction with the external medium, to allow for only mildly-relativistic sheath, to include the Kelvin-Helmholtz instability modes etc.  Such studies are under way.

\section*{Acknowledgements}

KT is partly supported by JSPS Grants-in-Aid for Scientific Reseach 15H05437 and also by a JST grant ``Building of Consrtia for the Development of Human Resources in Science and Technology''. SSK is supported via STFC grant ST/N000676/1. 
OP is supported by the ERC synergy grant "BlackHoleCam: Imaging the Event Horizon of Black Holes" (Grant No. 610058).
The computations were carried out on the cluster Draco in Tohoku University.










\bsp	
\label{lastpage}
\end{document}